\documentclass[a4paper,twoside,twocolumn,english,prl,aps,showpacs,floatfix,groupedaddress,superscriptaddress]{revtex4}
\pdfoutput=1
\usepackage{mathptmx}
\usepackage[]{fontenc}
\usepackage[latin9]{inputenc}
\usepackage{amsbsy}
\usepackage{graphicx}
\usepackage{amssymb}

\makeatletter

\newcommand{\binom}[2]{{#1 \choose #2}}

\@ifundefined{textcolor}{}
{%
 \definecolor{BLACK}{gray}{0}
 \definecolor{WHITE}{gray}{1}
 \definecolor{RED}{rgb}{1,0,0}
 \definecolor{GREEN}{rgb}{0,1,0}
 \definecolor{BLUE}{rgb}{0,0,1}
 \definecolor{CYAN}{cmyk}{1,0,0,0}
 \definecolor{MAGENTA}{cmyk}{0,1,0,0}
 \definecolor{YELLOW}{cmyk}{0,0,1,0}
 }

\usepackage{bm}

\makeatother

\usepackage{babel}

\begin{document}

\title{From Rotating Atomic Rings to Quantum Hall States}

\author{M. Roncaglia}

\affiliation{Dipartimento di Fisica del Politecnico, corso Duca degli Abruzzi
24, I-10129, Torino, Italy}

\affiliation{Max-Planck-Institut f\"ur Quantenoptik, Hans-Kopfermann-Str. 1,
D-85748, Garching, Germany}

\author{M. Rizzi}

\affiliation{Max-Planck-Institut f\"ur Quantenoptik, Hans-Kopfermann-Str. 1,
D-85748, Garching, Germany}

\author{J. Dalibard}

\affiliation{Laboratoire Kastler Brossel, CNRS, UPMC, \'Ecole
normale sup\'erieure, 24 rue Lhomond, 75005 Paris, France}

\begin{abstract}
Considerable efforts  are currently devoted to the preparation of ultracold neutral atoms in the emblematic strongly correlated quantum Hall regime. The routes followed so far essentially rely on thermodynamics, i.e. imposing the proper Hamiltonian and cooling the system towards its ground state. In rapidly rotating 2D harmonic traps the role of the transverse magnetic field is played by the angular velocity. For particle numbers significantly larger than unity, the required  angular momentum is very large and it can be obtained only for spinning frequencies extremely near to the deconfinement limit; consequently, the required control on experimental parameters turns out to be far too stringent. Here we propose to follow instead a dynamic path starting from the gas confined in a rotating ring. The large moment of inertia of the fluid facilitates the access to states with a large angular momentum, corresponding to a giant vortex. The initial ring-shaped trapping potential is then adiabatically transformed into a harmonic confinement, which brings the interacting atomic gas in the desired quantum Hall regime. We provide clear numerical evidence that for a relatively broad range of initial angular frequencies, the giant vortex state is adiabatically connected to the bosonic $\nu=1/2$ Laughlin state, and we discuss the scaling to many particles.
\end{abstract}

\maketitle
While coherence between atoms finds its realization in Bose--Einstein condensates \citep{Cornell95,Ketterle95,Bloch08}, quantum Hall states \citep{Laughlin83} are emblematic representatives of the strongly correlated regime. The fractional quantum Hall effect (FQHE) has been
discovered in the early 1980s by applying a transverse magnetic field to a two-dimensional (2D) electron gas confined in semiconductor heterojunctions \citep{Tsui82}. Since then, FQHE has never stopped to intrigue the scientific community due to non-trivial transport properties and exotic
topological quantum phases \citep{ezawa08}. Such interest has also influenced the research in ultracold atomic gases, which in the last decade have been successfully exploited as a highly controllable playground for quantum simulations of many-body physics \citep{Bloch08}. The large versatility of these setups allows one to confine atoms in 2D harmonic traps and to impose an effective magnetic field either by rapid rotation \citep{Fetter08,Cooper08} or by laser-induced geometric gauge potentials \citep{spielman09}. In principle, such opportunity should allow one to experimentally explore the bosonic version of QHE, even if unfortunately it has been hitherto elusive.
 
From a theoretical point of view, a variety of interesting ground states (GS's)  have been identified for Bose gases as a function of the effective magnetic field \citep{Fetter08,Cooper08}.
At zero field, i.e., without rotation, the particles undergo Bose--Einstein condensation 
\citep{Cornell95,Ketterle95} and the atomic ensemble is superfluid. Differently from a rigid body, a superfluid of $N$ particles reacts to rotation with the formation of  quantized vortices, whose number $N_{\phi}$ increases with the rotation frequency. At large filling factor $\nu=N/N_{\phi}\gtrsim10$ an ordered vortex lattice is formed. For $\nu<10$, the lattice melts because of quantum fluctuations, which signals the breakdown of the mean-field  description and the access into
the FQHE regime. The filling factor is now better defined as $\nu=N/m_{\mathrm{max}}$ with $m_{\mathrm{max}}$ the maximum angular momentum occupied by single particles. FQHE states are obtained for values of $\nu$ of order unity, which correspond to very large total angular momenta
$L\propto N^{2}$.

Like in solid-state physics, most of the preparation procedures employed so far in rotating atomic ensembles approached the GS by cooling down the system with a fixed Hamiltonian. By contrast we explore in this paper an alternative method that consists in starting from an easily preparable state (typically uncorrelated), following a dynamic route by changing an external parameter, and eventually obtaining the desired state. This strategy  has been successful for the experimental investigation of the superfluid to Mott insulator transition in optical lattices \citep{Greiner02}. We propose to implement it to reach Quantum Hall states with the following steps: (i) We engineer a Mexican-hat trapping potential by superposing a standard harmonic trap with the repulsive potential created by a ``plug" laser beam, which is focused at the center of the trap. (ii) By stirring the gas, we prepare the $N$ bosonic atoms in a giant vortex state, corresponding to the lowest energy state of the Mexican-hat potential for a given angular momentum $L$. (iii) The stirring is removed and the plug is adiabatically switched off. (iv) In the final harmonic trap, we obtain the GS with the initially imparted angular momentum $L$, thanks to rotational symmetry. We show that if $L=N(N-1)$ then the 2-body contact interactions drive the gas into the celebrated bosonic $\nu=1/2$ Laughlin state \citep{Laughlin83}.

\section*{Single particle physics}
\label{sec:single_particle_physics}

\begin{figure*}
\hspace{-6.5cm}\includegraphics[width=0.8\columnwidth]{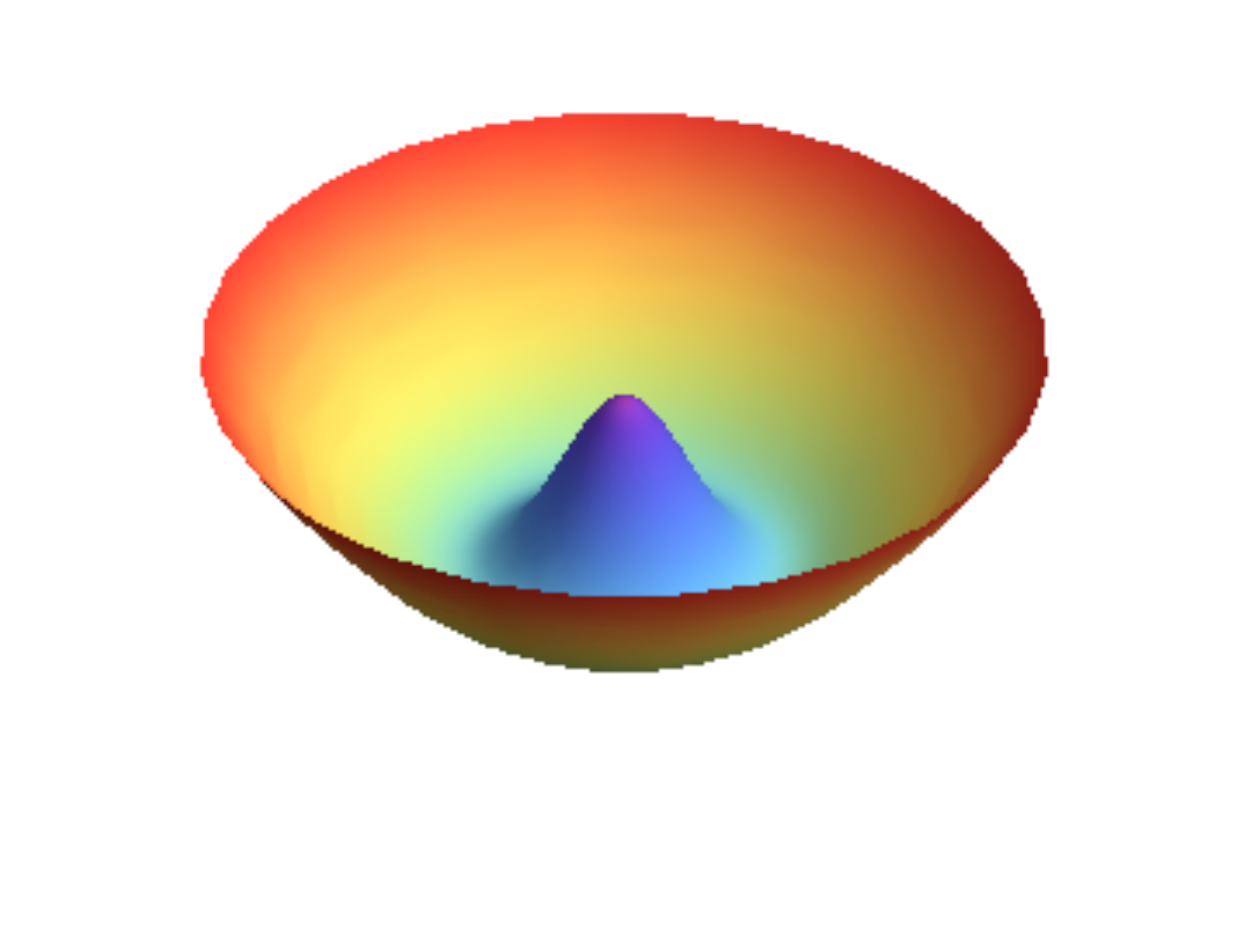}\hspace{-0.5cm}
{\includegraphics[width=0.7\columnwidth]{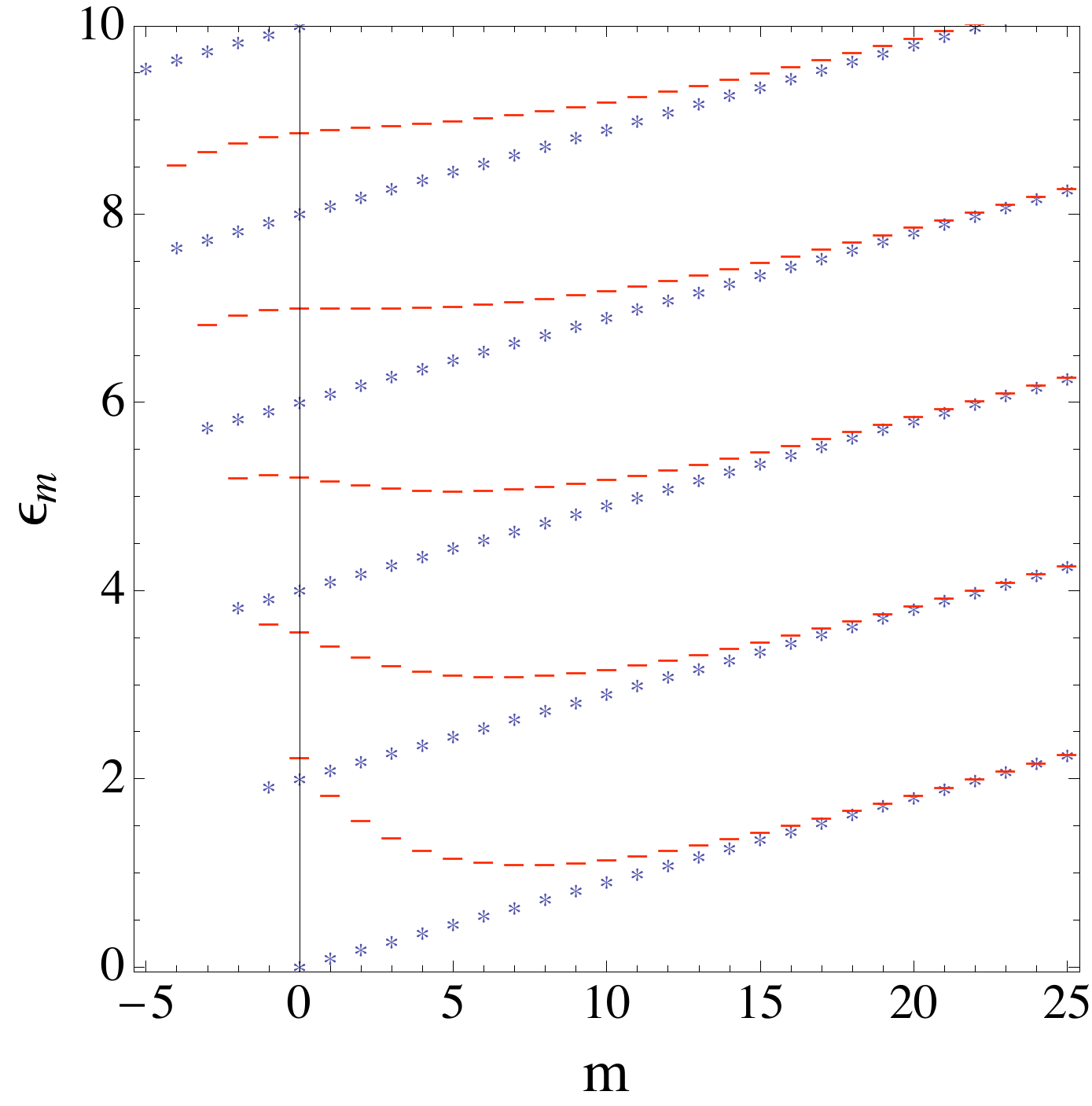}}
\put(20,40){\includegraphics[width=0.6\columnwidth]{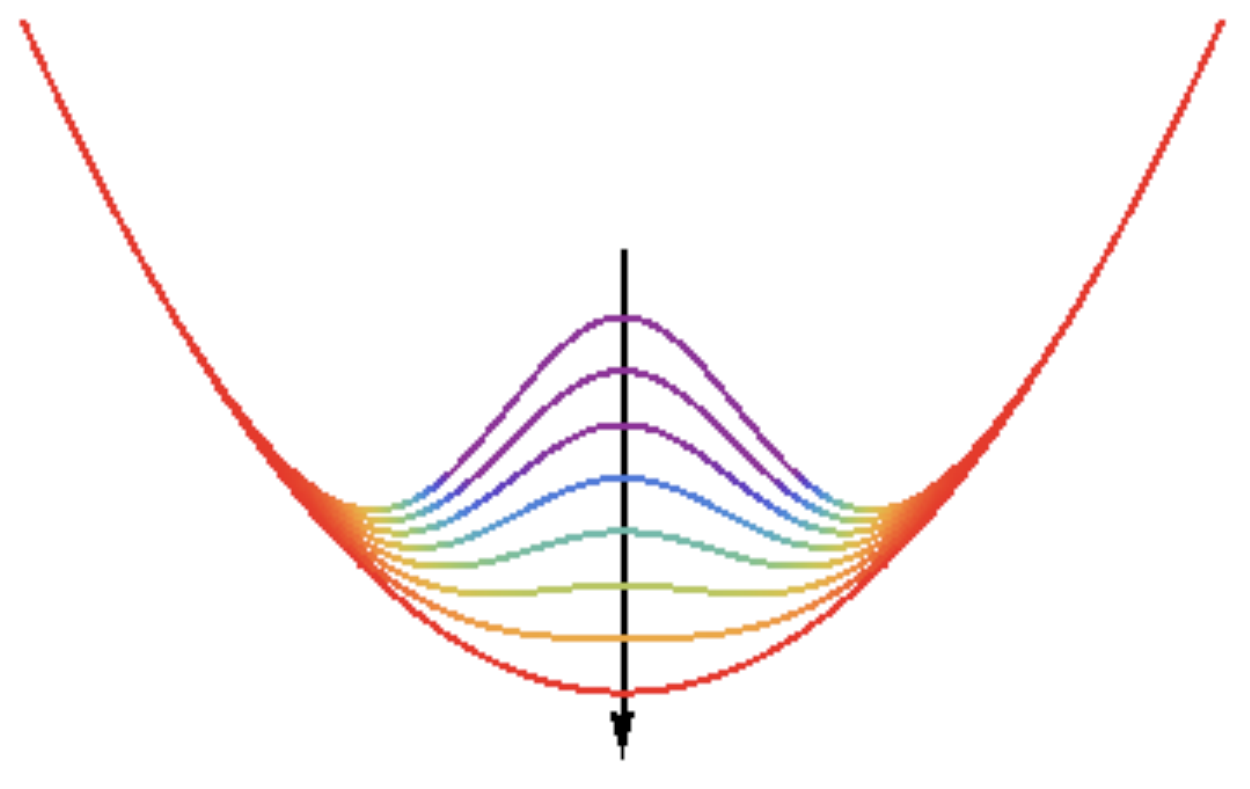}}
\par \hspace{-3.5cm} (a) \hspace{5cm}  (b) \hspace{6cm} (c)
\caption{\label{fig:potential3D}\textbf{\small Mexican hat potential:}{\small 
(a) Bosonic atoms are confined in a combined trap with (i) an isotropic harmonic confinement and (ii) the dipole potential created by a blue-detuned, gaussian laser beam that plugs the trap center and pushes the particles away from this point. The resulting potential exhibits a Mexican-hat shape. (b) Under fast rotation, the single-particle energy spectrum exhibits a Landau Level picture (red dashes), where levels are arranged in quasi-degenerate manifolds as in the case of a purely harmonic trapping (blue asterisks). The level plot has been drawn for $\delta=0.09$, $\alpha=3.0$ and $w^2=8.0$, parameters that we will use in the many-body problem for $N=9$. The single-particle minimum of energy can be adjusted to any desired value on angular momentum by tuning the waist and the power of the plug beam ($m=8$ in the present case). (c) In order to enter FQHE regime of the many-body interacting system, we propose to switch off the plug beam, eventually recovering the usual parabolic form.}}

\end{figure*}

In experiments with rotating atomic gases, particles are usually trapped by a harmonic potential and stirred by time-varying magnetic field or auxiliary laser beams  \citep{Fetter08}. 
In the frame rotating at angular speed $\boldsymbol{\Omega}=\Omega\mathbf{z}$, the Hamiltonian of a single particle in the harmonic trap of frequencies $(\omega,\omega,\omega_{z})$ can be written as 
 \begin{equation}
\mathcal{H}_{\mathrm{trap}}=\frac{1}{2M}(\mathbf{p}-\mathbf{A})^{2}+\frac{M}{2}(\omega^{2}-\Omega^{2})(x^{2}+y^{2})+\frac{M}{2}\omega_{z}^{2}z^{2},\label{eq:Htrap} 
\end{equation}
with $\mathbf{A}=M\boldsymbol{\Omega}\times\mathbf{r}=M\Omega(-y,x,0)$. 
In the following we suppose that all relevant energies are much smaller than $\hbar \omega_z$, so that the motion along the $z$ direction is  frozen and the problem is effectively two-dimensional. 
In the limit of centrifugal deconfinement $\Omega\to\omega$,  the system is formally equivalent to bosons of charge $q=1$ in uniform magnetic field $\mathbf{B}=\boldsymbol{\nabla}\times\mathbf{A}=2M\Omega\,\mathbf{z}$.
From now on, we express energies and lengths in units
of $\hbar\omega$ and $\sqrt{\hbar/M\omega}$, respectively. It is
well known \citep{Fetter08} that the problem can be rewritten as
two decoupled harmonic oscillators $\mathcal{H}_{\mathrm{trap}}=(2a^{\dagger}a+1)+\delta(b^{\dagger}b-a^{\dagger}a)$,
in terms of ladder operators $a$, $b$, and $\delta=1-\Omega/\omega$
is the frequency offset. Every state is labeled by the occupation
number $n_{a}$, $n_{b}$ of the two modes, and it is denoted as $|\psi_{n_b,n_a}\rangle$.
Note that a gauge field similar to the one entering into (\ref{eq:Htrap}) can also be induced by  geometric phases instead of rotation \cite{spielman09}. The scheme outlined in the present paper should work equally well in this case, the only significant difference being that (\ref{eq:Htrap}) is now the single-particle Hamiltonian in the laboratory frame, instead of the rotating frame. 

In the limit $\delta\ll1$, the quantum number $n_{a}$ identifies different manifolds called Landau Levels (LL). Within each LL, the states (labeled by $n_{b}$) are quasi-degenerate due to the small separation energy $\delta$. The quantity $m=n_{b}-n_{a}$ is the angular momentum of
the particle. In the lowest Landau level (LLL), $n_{a}=0$ and the one-body eigenfunctions assume the simple expression $\psi_{m,0}(z)=\frac{1}{\sqrt{\pi m!}}z^{m}e^{-\left|z\right|^{2}/2}$, where $z$ now denotes the position in the complex plane ($z=x+iy)$, with energies $E_{m}=m\delta$ and angular momentum $m$.

The first key feature of our proposition is to replace the ordinary harmonic potential with a Mexican-hat one, like in Fig.\ref{fig:potential3D}. This can be done by shining the center of the  harmonic trap with a laser beam prepared in a circular, Gaussian TEM$_{00}$ mode \citep{bretin04}. When the laser frequency is chosen larger than the atomic resonance frequency (`blue detuning'), the laser beam creates a repulsive dipole potential proportional to the light intensity. The beam is chosen to be perpendicular to the $xy$ plane and the dipole potential is of the form
 \begin{equation}
U_{w}(x,y)=\alpha\exp\left[-{2(x^{2}+y^{2})}/{w^{2}}\right],\label{eq:Ur}
 \end{equation}
where $w$ is the laser waist and $\alpha$ is proportional to the laser intensity. The sum of the harmonic potential $(x^2+y^2)/2$ and of $U_w(x,y)$ has a bump in $x=y=0$ in the laboratory frame  when $\alpha>w^{2}/4$. 

At moderate intensities of the plug, as the ones employed in our preparation scheme, the classification of single-particle energy eigenstates in terms of LL remains valid (see Fig. \ref{fig:potential3D} and Methods). In the LLL the single-body energies are in good approximation:
 \begin{equation}
\epsilon_{m}=m\delta+\alpha\left(1+\frac{2}{w^{2}}\right)^{-(m+1)}.\label{eq:E_m} 
\end{equation}
At fixed laser parameters $\alpha$ and $w$, the angular momentum $m$ that minimizes $\epsilon_{m}$ is a decreasing function of the rotation frequency offset $\delta$. We denote by $\delta_m$ the value for which the level crossing $\epsilon_{m+1}=\epsilon_{m}$ occurs. The LLL state with angular momentum $m$ is thus the lowest energy state when $\delta$ is chosen in the interval  $\delta_{m}<\delta<\delta_{m-1}$,
whose width is
 \[
I_m=\delta_{m-1}-\delta_{m}=\alpha\left(\frac{2}{w^{2}}\right)^{2}\left(1+\frac{2}{w^{2}}\right)^{-(m+2)}.
 \]
Later on, we will be interested in choosing a specific value $m=\ell$ and in maximizing the width $I_\ell$ of the stability window. This can be done, at fixed intensity $\alpha$, by choosing $w^{2}=\ell$. 
The central rotation frequency in the stability window for $\ell$ then corresponds to 
 \begin{equation}
\delta_{\ell}^{c}=\frac{1}{2}(\delta_{\ell-1}+\delta_{\ell})=2\alpha\frac{\ell+1}{\ell^{2}}\left(1+\frac{2}{\ell}\right)^{-(\ell+2)}.\label{eq:domega center}
\end{equation}
Notice that for large values of $\ell$, we get $\delta_{\ell}^{c}\propto\alpha\ell^{-1}$,
thus if we want to keep it sizable, we have to choose  $\alpha\propto\ell$.

\section*{Many-body physics}

In the context of cold bosonic gases in the LLL subspace, two-particles interactions can
be modelled by the contact potential 
 \begin{equation}
\mathcal{H}_{\mathrm{2}}=c_{2}\sum_{i<j}\delta^{(2)}(z_{i}-z_{j}),\label{eq:2-body delta potential}
\end{equation}
whose strength is given by the adimensional parameter $c_{2}=\sqrt{8\pi}a_{s}/a_{z}$,
where $a_{s}$ is the 3D $s$-wave scattering length  and $a_{z}=\sqrt{\hbar/M\omega_{z}}$
is the size of the ground state in the strongly confined direction \citep{Fetter08,Cooper08}. Within the kernel of
$\mathcal{H}_{\mathrm{2}}$ the $\nu=1/2$ Laughlin state\begin{equation}
\Psi_{\mathrm{Lau}}=\left[\prod_{i<j}(z_{i}-z_{j})^{2}\right]e^{-\sum_{j}|z_{j}|^{2}/2}\label{eq:laughlin}\end{equation}
has the lowest total angular momentum $L_{\mathrm{Lau}}=N(N-1)$, or equivalently the angular momentum per particle $\ell_{\mathrm{Lau}}=N-1$. 

\begin{figure}
\centering
\hspace{2.5cm}
{\includegraphics[width=0.68\columnwidth]{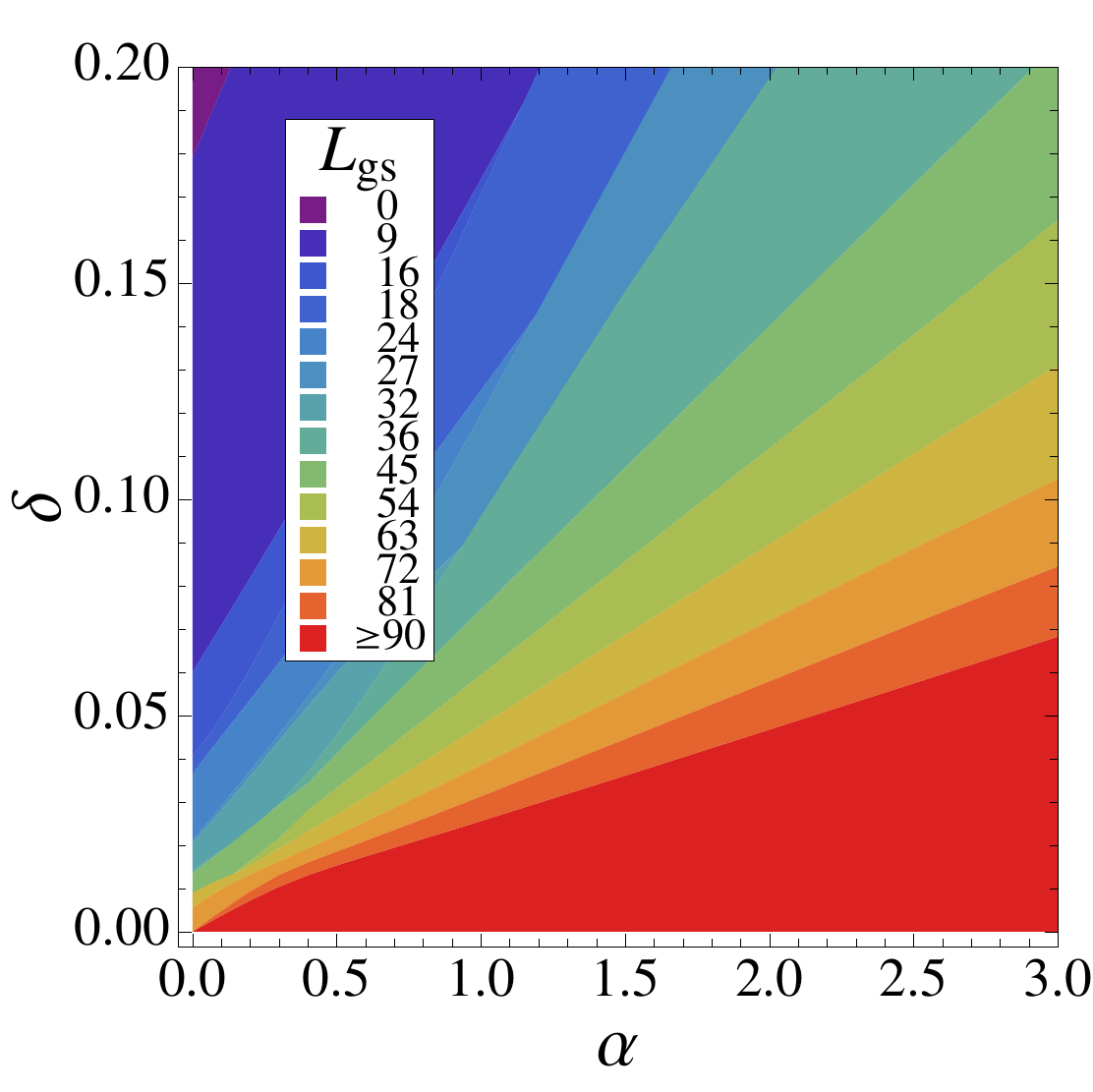}}
\put(-240,10){\includegraphics[width=0.23\columnwidth]{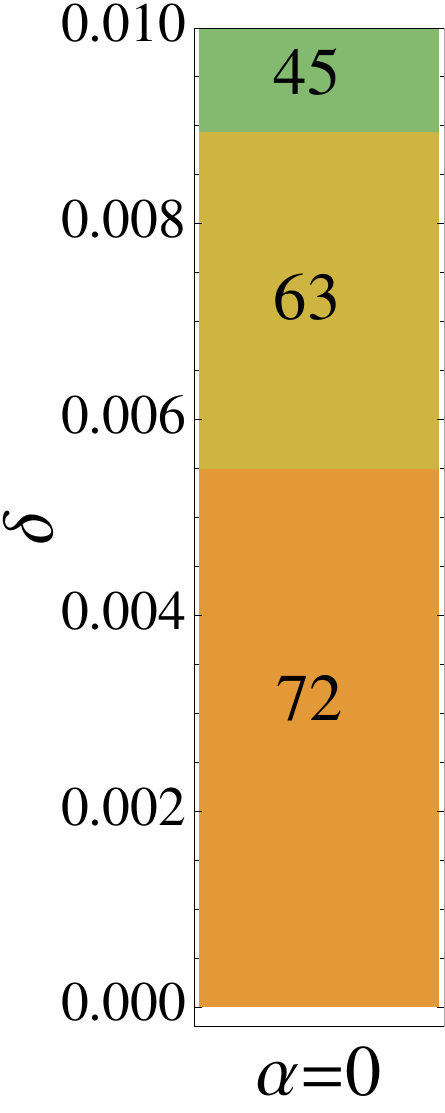}}
\par \vspace{-0.4cm}\hspace{-4.5cm} (a) \hspace{2cm}  (b) 
\caption{\textbf{\label{fig:Rainbow}}\textbf{\small 
Phase diagram for $\mathbf{N=9}$ and $\mathbf{c_2=0.5}$.}{\small{}
Exact diagonalization in a truncated LLL basis ($m\le2N=18$) is performed
separately for each sector of $L$, thanks to rotational invariance.
Energies are then shifted by the total angular momentum term $L\, \delta$
to draw the phase boundaries. (a) In absence of the plug beam, $\alpha=0$,
the window of stability for Laughlin angular momentum ($L_{\mathrm{Lau}}=72$)
is narrow and extremely close to deconfinement limit. (b) Conversely,
the region with $L_{\mathrm{Lau}}$ opens up and drifts away from $\delta=0$ as the plug power $\alpha$ is ramped up (at constant $w^2=8$); the same happens for other large angular momenta around
it. For large values of $\alpha$ (typically larger than unity), the GS found when varying $\delta$ are essentially non-correlated states, where all atoms accumulate in the same giant vortex state. Consequently the total angular momentum $L$ undergoes jumps of size $N$, corresponding to the addition of one flux quantum to each particle.}}
\end{figure}

We first recall the practical difficulties to attain the FQHE regime via a thermodynamic route for a pure harmonic confinement in the $xy$ plane. In a typical experiment with $^{87}$Rb atoms ($a_{s}= 5\,$nm), a longitudinal frequency $\omega_{z}/2\pi=50\,$kHz gives $a_z=50\,$nm and an interaction parameter $c_{2}=0.5$. Then, already for a modest number of particles, the Laughlin state is reached only  for rotation frequencies $\Omega$ extremely close to the centrifugal limit $\omega$. For $N=9$ 
and $c_2=0.5$, we find using exact numerical diagonalization that the Laughlin state is the GS only for $\delta_{\mathrm{Lau}}< 5.5\;10^{-3}$ (see figure \ref{fig:Rainbow}a). 
This very low threshold makes it difficult to transfer the desired angular momentum to the gas. 
Indeed  when the stirrer consists in a \emph{rotating} anisotropic potential $\epsilon(x^2-y^2)/2$, the corresponding anisotropy $\epsilon$ must be chosen smaller than $\delta$ to avoid a dynamical instability of the center-of-mass motion \citep{Rosenbusch02}.  One has thus to restrict to extremely weak stirrers, with $\epsilon$ in the $10^{-3}$ range. However one must also choose $\epsilon\gg u$, where $u$ is the strength of the \emph{static} anisotropy defect $u(x^2-y^2)/2$, otherwise the gas cannot  be effectively set in rotation. Unfortunately, in realistic traps the typical values of $u$ are at least of the order of $10^{-3}$.  Consequently it is quite problematic to fulfil simultaneously these various constraints. Lastly, we mention that in such a thermodynamical procedure the temperature has to be kept below $\delta$, i.e the gap between the Laughlin state and the GS with $L=L_\mathrm{Lau}+1$ (the interaction energy is zero in both cases), which is a very stringent requirement.

Here we propose a different point of view where the FQHE regime can be tackled from a dynamical perspective, with a two-step procedure. The first step is addressed in this section and it consists in the preparation of a giant vortex state of $N$ particles in the Mexican-hat potential of Fig.  \ref{fig:potential3D}a, with the desired angular momentum $L=L_{\mathrm{Lau}}$. This preparation is easier than the direct production of the Laughlin state, thanks to the favourable parameter sensitivity of the Mexican-hat potential. More precisely the increased moment of inertia of the gas enables one to reach $L\propto N^{2}$ in a relatively broad interval of $\Omega$. The second step involves the adiabatic transformation of the giant vortex state in the Laughlin state, and it will be analyzed in the next section.

In the limit case of no interactions, every single particle in the Mexican hat potential should rotate at angular momentum $\ell=L/N$. The bosons then condense in the giant vortex state
 \begin{equation}
 \Psi_{\mathrm{v}}^{(N)}(\ell)=\left[\prod_{i=1}^{N}z_{i}^{\ell}\right]e^{-\sum_{j}|z_{j}|^{2}/2},   
 \label{eq:giant vortex}
 \end{equation}
similar to those already observed in \citep{Ryu07} and theoretically analyzed in  \citep{Kasamatsu02}. As shown in the previous section, the window of stability $I[\ell_{\mathrm{Lau}}]$ is optimized for $w^{2}=\ell_{\mathrm{Lau}}$. For $N=9$ the constraint $\alpha>w^2/4$ imposes $\alpha>2$. We choose in the following $\alpha=3$, which leads to $\delta_{\mathrm{Lau}}\in(0.081,\,0.101)$.

In the presence of interactions, the interval of values for $\delta$ leading to a ground state with $L=L_{\rm Lau}$ can be determined either from a Bogoliubov analysis or from exact diagonalization. The main role of the interactions is to deplete the contribution of the mean angular momentum $\ell$ in favour of neighbouring ones $\ell \pm q$, with $q\ll \ell$. The Bogoliubov analysis is well suited for strong plugs ($\alpha>1$) where the depletion is small, while for intermediate regimes a full many-body numerical treatment is needed (see Methods). Thanks to the angular momentum conservation, the exact diagonalization can be performed in each $L$ sector separately and the conjugate variable $\delta$ simply yields the energy shifts $L\delta$. The phase diagram as a function of $(\alpha$, $\delta)$ is presented in Fig.\ref{fig:Rainbow}b for $N=9$, $c_2=0.5$. It strongly supports the  sketch drawn before for non-interacting particles. In particular the computed ground state for a Mexican-hat potential with $\alpha=3$ possesses the required angular momentum $L_{\mathrm{Lau}}$ for the interval 
$\delta_{\mathrm{Lau}} \in (0.084,\,0.105)$, very close to the one in absence of interaction. This corresponds to a $\sim 10\%$ frequency difference between $\Omega$ and $\omega$, which is notably larger than the typical stirrer anisotropy needed to set a gas in rotation. This ensures that the preparation of the giant vortex state with $L=L_{\rm Lau}$ should be rather robust.

It is important to stress that the Mexican-hat potential  is employed just for the scope of injecting the right quantity of angular momentum $L_{\mathrm{Lau}}$, and not for producing the Laughlin state itself. The situation considered here is thus completely different from former proposals suggesting to find a tradeoff between $\Omega$ and $\alpha$ that optimizes the fidelity with the Laughlin state \citep{morris07}.

\section*{Adiabatic evolution }

Once the gas has gained the desired angular momentum $L=N(N-1)$ via equilibrating
in the giant vortex state (\ref{eq:giant vortex}), the stirrer at frequency $\Omega$ can be suppressed. The situation becomes rotationally symmetric and the total angular momentum is thus conserved. Then, the slow removal of the laser plug will result in a redistribution of particles around the mean angular momentum $\ell_{\mathrm{Lau}}$  by repulsive interactions. Such a redistribution reaches a paradigmatic form in the unplugged harmonic trap, where the Laughlin state (\ref{eq:laughlin}) has no interaction energy anymore. From a technical point of view, we note that in absence of stirring we can now look for the GS of the gas in the laboratory frame, within the subspace of the $L_z$ that had been imparted to the cloud during the stirring phase.
\begin{figure}
{\hspace{-7cm}  (a)}
\par \vspace{-0.4cm}  \includegraphics[width=0.74\columnwidth]{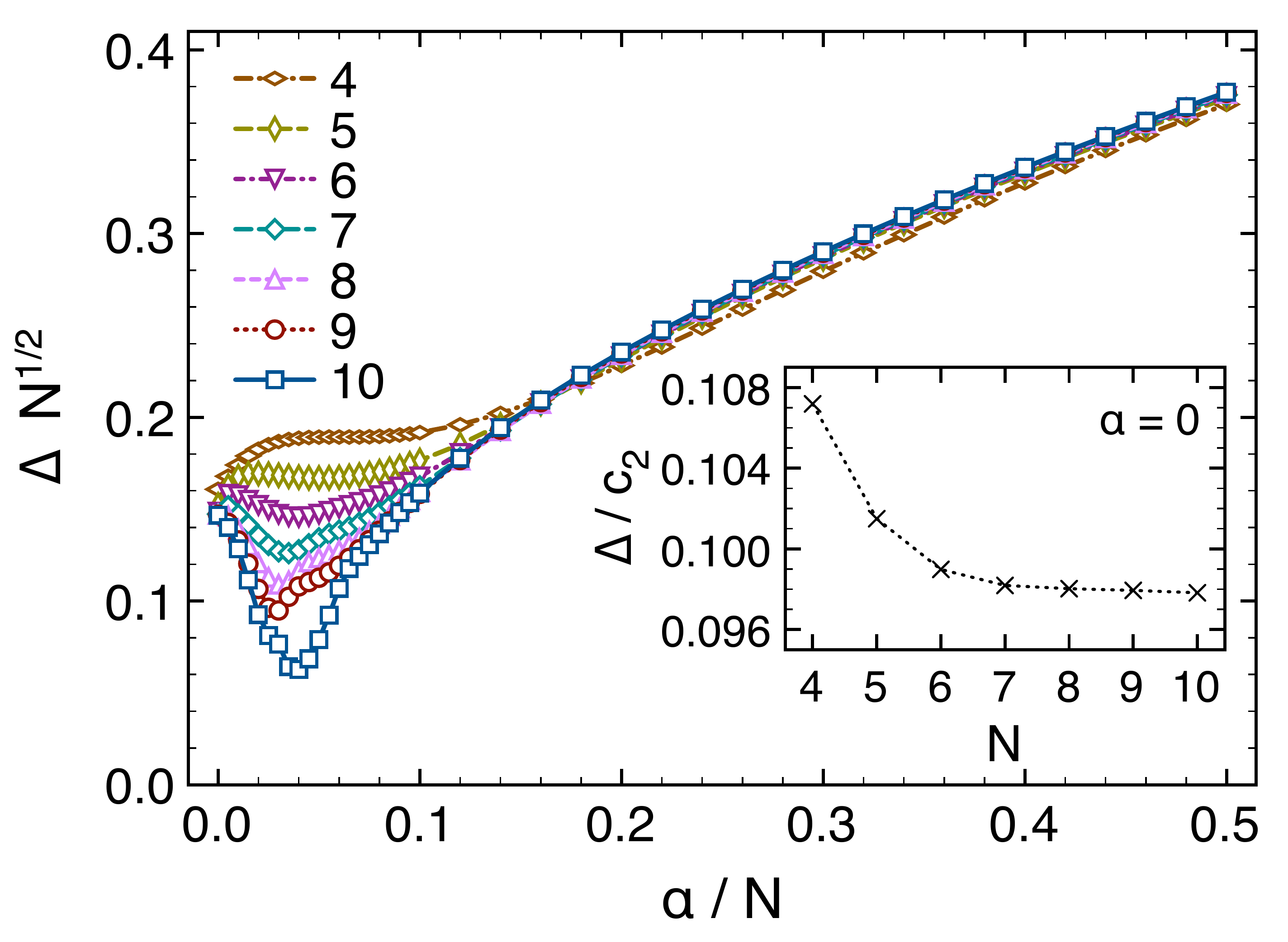}\par
{\hspace{-7cm}  (b)}
\par \vspace{-0.4cm}  \includegraphics[width=0.74\columnwidth]{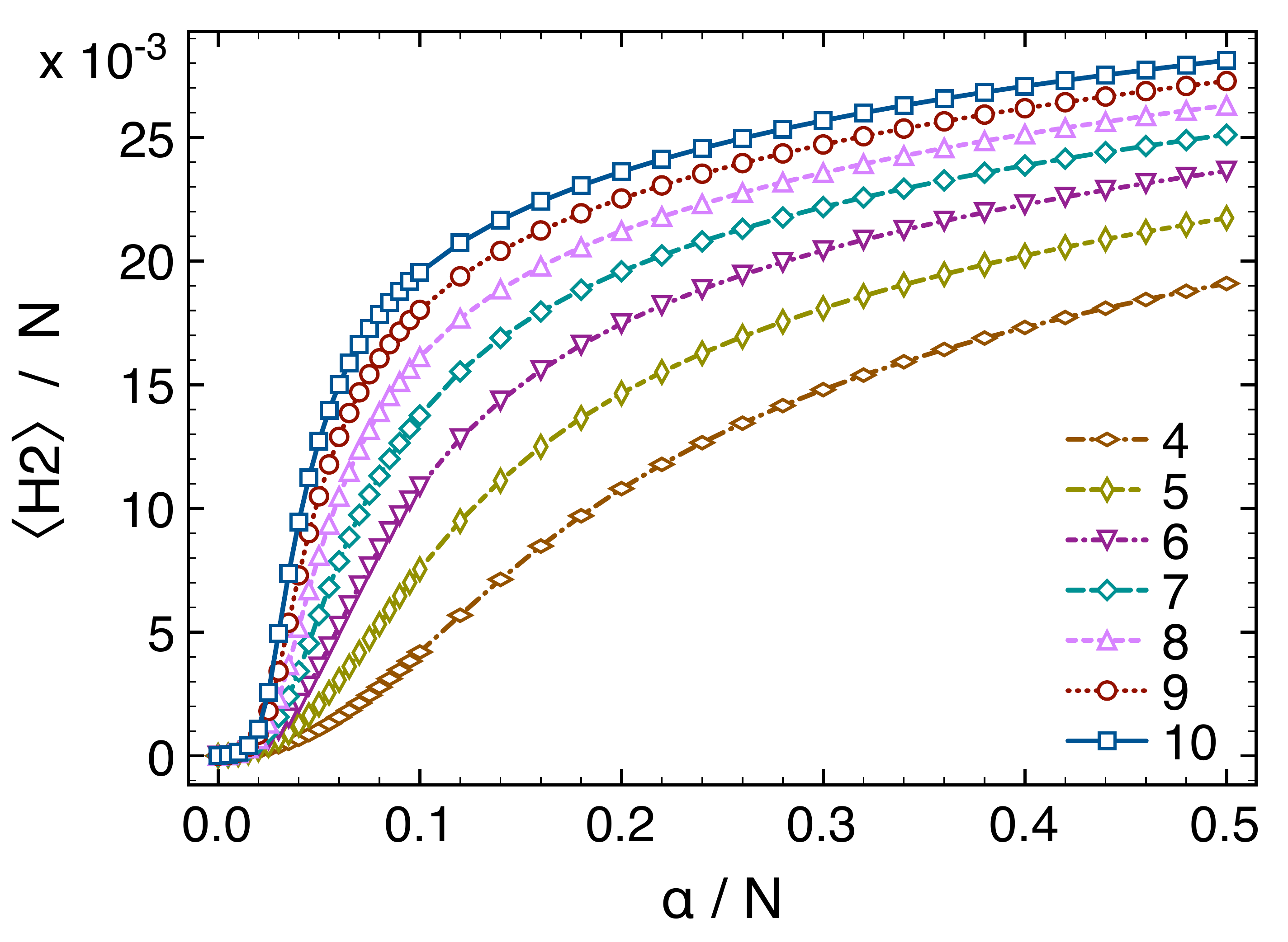}\par
{\hspace{-7cm}  (c)}
\par \vspace{-0.4cm}  \includegraphics[width=0.74\columnwidth]{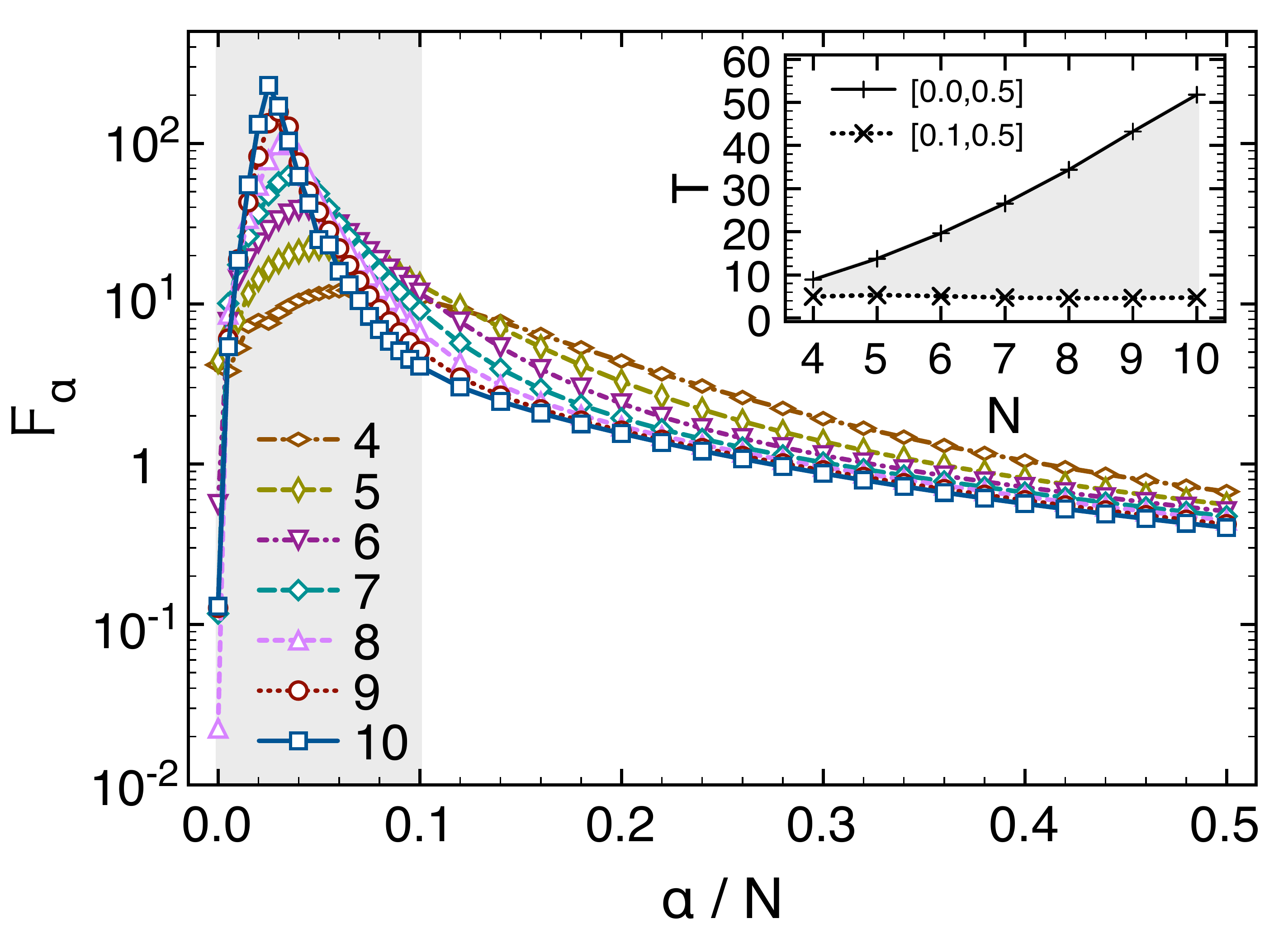}
\caption{\textbf{\label{fig:Finite-size-scaling}}
\textbf{\small Finite size scaling analysis.}{\small{}
Data collapse of (a) the gap and (b) interaction energy
$\langle\mathcal{H}_{2}\rangle$ as a function of $\alpha$ for an initial value 
of $\mu=c_2 N^{1/2}/(2\pi^{3/2})\sim 0.15$, corresponding to $c_{2}=0.5$ for $N=9$. The inset in plot (a) shows the finite size scaling of the Laughlin gap in the harmonic case ($\alpha=0$). Our estimates gives $\Delta_\mathrm{Lau}=0.097(1)$ in the thermodynamical limit.
(c) Scaling of the function 
$F_{\alpha}\equiv\Delta^{-2}\left|\left\langle \Psi_{0}\right|\left(\partial U_{w}/\partial\alpha\right)\left|\Psi_{1}\right\rangle \right|$
whose integral provides an estimation for the adiabatic time (inset).
At sizable enough $\alpha$'s ($\gtrsim0.1N$) the curves collapse
from above, giving a total time $T$ which is approximately constant
with $N$ ("$\times$" points). Alas the presence of a pronounced bump for small 
$\alpha$'s leads to $T\propto N$ ("$+$" points). Alternative
strategies that lead to lower adiabatic times are discussed in Fig.
\ref{fig:adiabaticity}.}}
\end{figure}

The system will follow the instantaneous GS $\Psi_0$ if the unplugging path can be followed slowly enough to satisfy the adiabatic condition $\left|\left\langle \Psi_{0}\right|  \left(\partial\mathcal{H}/\partial t\right)\left|\Psi_{j}\right\rangle \right|\ll\Delta_{j}^{2}$, where $\Psi_j$ represents an excited eigenstate of energy $E_j$ of the instantaneous Hamiltonian $\mathcal{H}$, and where $\Delta_j=E_{j}-E_{0}$ \citep{Messiah61}.  We have checked numerically that the most stringent constraint originates from the first excited state $\Psi_1$ and we thus focus our discussion on this state. Once the gap $\Delta=E_{1}-E_{0}$ is known from exact diagonalization, the changing rate of laser intensity $\alpha$ and/or its rescaled cross section $\sigma\equiv w^2/(N-1)$ with time $t$ is determined by the condition
\begin{equation}
F_{\alpha}\mathrm{d}\alpha+F_{\sigma}\mathrm{d}\sigma\ll\mathrm{d}t\ ,\label{eq:adiabcond}
\end{equation}
where $F_{x}\equiv\Delta^{-2}\left|\left\langle \Psi_{0}\right|\left(\partial U_{w}/\partial x\right)\left|\Psi_{1}\right\rangle \right|$
is the matrix element of the potential variation in $x=\alpha,\sigma$. The minimal total time $T$ required for adiabaticity will then be the integral of those functions along the chosen path, $T=\int\left(F_{\alpha}\mathrm{d}\alpha+F_{\sigma}\mathrm{d}\sigma\right)$. In the following, we first consider the case where the waist is kept fixed, which is experimentally straightforward since it involves only a variation of the laser intensity; then we address the general case of changing of both $\alpha$ and $\sigma$.

We have performed numerical simulations for up to $N=10$ particles, in a LLL truncated single-particle basis $m\le2N$, in order to test the validity of the adiabatic approximation  (see Methods). For the chosen test case of $N=9$, $c_{2}=0.5$, ramping down the intensity from the initial value $\alpha=4.5$ at constant $\sigma=1$, we obtain $T\approx 43$ (in units of $\omega^{-1}$). Such a value of $T$ is a reasonable time in state-of-art experiments, establishing the feasibility of our scheme for $N=9$, as opposed to the procedure involving a purely harmonic rotating trap. 

The exponentially increasing dimension of the Hilbert space and the strong correlations involved ward off going much further than $N=10$ particles with the exact diagonalization method. To infer the behavior of larger samples, we performed finite size scaling of the relevant energies using the Bogoliubov approximation (see Methods). Our scheme requires the preparation of the gas in the ring with $\alpha\propto N$ and $\sigma=1$, for which the chemical potential goes as $\mu\simeq c_{2}N^{1/2}/(2\pi^{3/2})$.  The LLL approximation requires $\mu<2$, and working at fixed $\mu$  implies $c_{2}\propto N^{-1/2}$. We then deduce  that the energy gap behaves like $\Delta\propto N^{-1/2}$ and the interaction energy as $\langle\mathcal{H}_{2}\rangle\propto N$ (see Methods). We have plotted in Fig.\ref{fig:Finite-size-scaling}(a)-(b) the variations of the gap $\Delta$ and the interaction energy $\langle \mathcal{H}_2\rangle$ using $c_{2}=1.5 N^{-1/2}$. The expected data collapse is well verified for  values of $\alpha/N$ larger than 0.1. 

A finite-size scaling can be performed also for the quantity $F_\alpha$ entering the adiabatic condition  (\ref{eq:adiabcond}) and we plot the result at fixed waist $\sigma$ in  Fig.\ref{fig:Finite-size-scaling}c. This function takes its largest values in the interval $\alpha \in (0,\alpha_c)$ with $\alpha_c=0.1/N$. We can therefore decompose the adiabatic path into two successive parts. In the first part the plug laser intensity $\alpha$ is decreased from  $\alpha_i=0.5/N$ down to $\alpha_c$ and this can be down adiabatically in a relatively short time $T_1=\int_{\alpha_c}^{\alpha_i}F_\alpha \, d\alpha\sim 5$, independent of the number of particles (inset of Fig.\ref{fig:Finite-size-scaling}c). In the second part ($0\leq \alpha\leq \alpha_c$) the breakdown of the scaling $\Delta \propto N^{-1/2}$ imposes a slowdown in the reduction of the plug intensity. The time $T_2=\int_{0}^{\alpha_c}F_\alpha \, d\alpha$ needed for this second part actually show a linear increase with $N$, hampering the feasibility for more than a few tens of bosons. 

\begin{figure*}[t]
\begin{centering}
\textbf{\includegraphics[width=2\columnwidth]{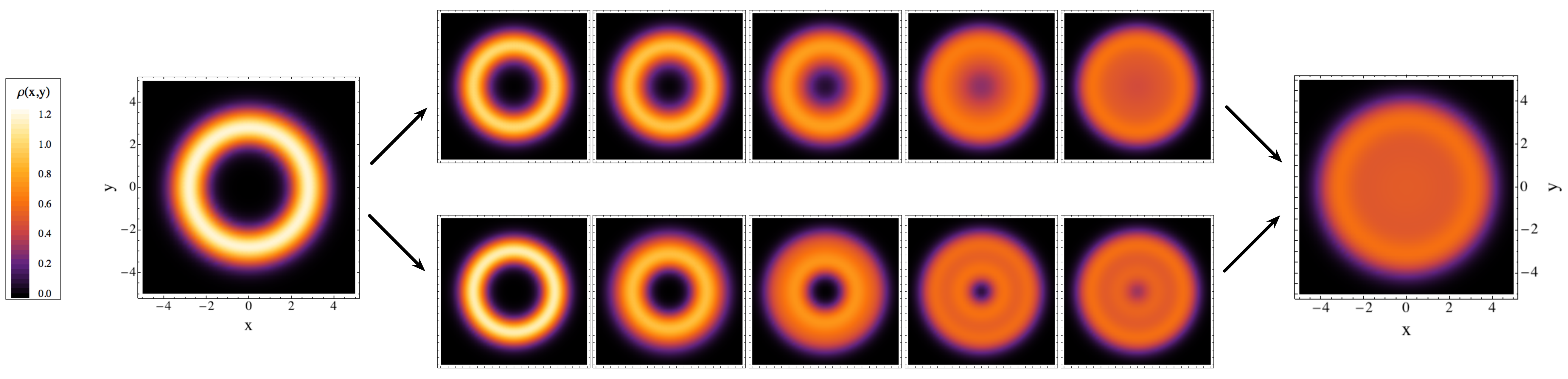}}
\par\end{centering}
\caption{\textbf{\label{fig:Density-profile}}\textbf{\small Density profile during adiabatic evolution ($\mathbf{N=9}$).}{\small{}  The leftmost panel corresponds to a giant vortex like structure, whereas the rightmost one depicts the flat disk shaped profile of the Laughlin state. In the upper row $\sigma=1$ is kept constant while $\alpha=1.,\,0.5,\,0.4,\,0.3,\,0.2,\,0.1,\,0.$ The last part of the ramp down procedure $0< \alpha\lesssim 0.1$ is the slowest, due to the large value of $F_{\alpha}$ in this region (see Fig. \ref{fig:Finite-size-scaling}c). In the lower row we squeeze the laser waist $\sigma=1.,\,0.5,\,0.25,\,0.125,\,0.025,\,0.00625,\,0.$ at fixed intensity $\alpha=1.$: particles spread towards the inner part of the trap in a different way, corresponding in a lower value of $F_{\sigma}$ and faster allowed rates of change. For systems within LLL, density profiles after trap release and time-of-flight imaging will simply display rescalings of these pictures. }}
\end{figure*}

\begin{figure}
\centering{}\includegraphics[width=0.9\columnwidth]{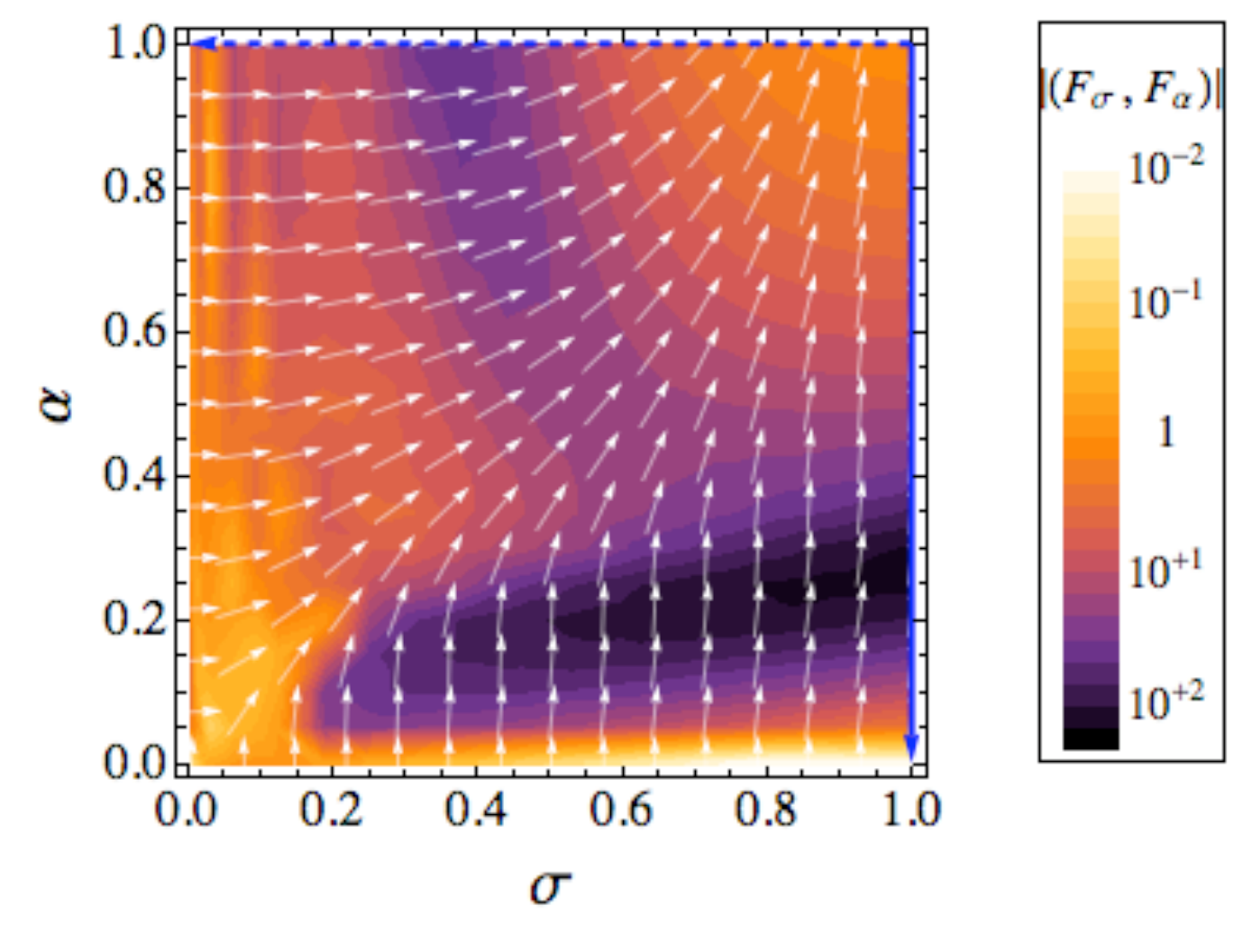}
\caption{\textbf{\label{fig:adiabaticity}}\textbf{\small Map of adiabaticity
requirements.}{\small{} Absolute value of the vector $(F_{\sigma},\, F_{\alpha})$ is plotted in the coloured map for $N=9$, evidencing the large value of $F_\alpha$ at large $\sigma=w^2/(N-1)$ and small $\alpha$, as well as the more favorable condition if one uses a reduction in time of the beam waist. The two paths described in the text give $T\sim40$ (solid blue line) and $T\sim20$ (dashed blue line). Superimposed white arrows represent the directions of the vector $( F_{\sigma}, F_{\alpha} )$. This plot  can serve for conceiving more intricate paths with the help of optimization techniques.}}
\end{figure}

The sequence of spatial density profiles during the time evolution is depicted in Fig.\ref{fig:Density-profile}a. From such a sequence it is clear that the gas starts feeding the trap center at the end of the path. An  alternative strategy to ramp down the plug consists on reducing its waist $w$ while maintaining $\alpha$ constant, this being performed in practice (up to the diffraction limit) using a motorized focusing optical element. The corresponding evolution of the density profile in the trap is represented in Fig.\ref{fig:Density-profile}b and in our specific case it can be covered adiabatically in half time with respect to the above situation (see Fig.\ref{fig:adiabaticity}).

A natural extension of our analysis is to consider a simultaneous ramping of $\alpha$ and $\sigma$, in order to minimize the total evolution time while fulfilling the adiabaticity criterion. To this aim, constrained optimization techniques can be implemented using the data of the vector $(F_{\sigma},F_{\alpha})$, represented in Fig.\ref{fig:adiabaticity}.  Experimentally, another  effective way of reducing the adiabatic ramp time is to increase the interaction coupling constant $c_{2}$, hence the gap,  via either Feshbach resonances \cite{Chin:2010} or a tighter longitudinal confinement $\omega_{z}$. For a ramp of $\alpha$ only, our numerical calculations with $N=9$  give $T\approx 65, 43, 20$ for $c_{2}=0.33, 0.5, 1.0$, respectively, corresponding to the empirical scaling law $T\approx 20\,c_2^{-1}$.

Finally we briefly address the consequences of some of the unavoidable experimental imperfections on the proposed scheme. The two principal perturbations that we can foresee are the imperfect centering of the plug beam and the residual trap anisotropy. We model these defects by writing the dipole potential created by the plug beam as $U_{w}^{'}=\alpha\exp\left[-2[(x-v)^{2}+y^{2}]/w^{2}\right]$ , and by adding the term $u(x^{2}-y^{2})/2$ to the single-particle Hamiltonian to account for the static anistropic defect. Here $v$ and $u$ are  dimensionless coefficients characterising these imperfections.  These two coupling terms break the rotation symmetry: in their presence, the angular momentum is not a conserved quantity anymore and the gas will undergo a cascade from $L=L_{\rm Lau}$ down to states with no angular momentum, by populating the first excited LL. To get a conservative estimate, we impose  the very stringent condition that the total angular momentum remains unchanged over the adiabatic ramp time, and we estimate the corresponding constraint on $u$ and $v$ using time-dependent perturbation theory  (see Methods). The constraint on $u$ is certainly the most challenging one. We find that the maximal tolerable trap anisotropy 
$u_{\rm max}\lessapprox 2\Delta_\mathrm{Lau}/N\approx 0.2\,c_{2}/N$. 
Taking $u\sim 10^{-3}$ as a realistic trap anisotropy,  we find that our scheme should be operational for atom numbers up to $N_\mathrm{max}=100$ for $c_2=0.5$.

\section*{Detection}

One of the simplest techniques to probe cold atomic setups consists in taking time-of-flight (TOF) pictures \citep{Bloch08}. The absorption image of the density profile expanded after releasing the harmonic confinement contains indeed useful informations about the initial situation in the trap. In the specific case of bosons in the LLL regime, the density profile is self-similar in time and the TOF picture simply magnifies the original particle distribution in the trap \citep{read03}. Given the direct connection between single-particle angular momenta and orbital radius (see Methods), a TOF image allows one to compute the angular momentum. The $\nu=1/2$ Laughlin state with $N$ particles exhibits a fairly flat profile of density $0.5$ inside a rim of radius $\sim \sqrt{N}$. Observing such TOF images would be already a first hint that one has effectively reached the QHE regime. 

Multi-particle correlations offer even more insight into the many-body state. These correlations are directly accessible if one uses a detection scheme that can resolve individual atoms with sub-micron resolution \citep{Bakr09,Sherson10}.
Alternatively the two-body correlation function can be tested at short distances using the resonant photo-association of spatially close pairs \citep{Gemelke10}.  The amount of produced molecules is indeed directly related to the correlation function $g^{(2)}(0)$, which is also in direct correspondence with the interaction energy $\langle\mathcal{H}_{2}\rangle/c_{2}$ (Fig.\ref{fig:Finite-size-scaling}b). Since the Laughlin state belongs to the kernel of $\mathcal{H}_{2}$, its presence will be signaled by a strong suppression of two-body losses. Moreover, in a strict analogy with solid state physics, we can imagine an experiment to measure FQHE plateaus in physical quantities. Namely, by varying the rotational offset $\delta$ in the giant vortex preparation stage it is possible to change $L$ by steps of $N$, i.e. move the penetrating magnetic flux in units of single quanta. Removing now the plug, the system will fall in a sequence of incompressible FQHE states: the final $g^{(2)}(0)$ is expected to display plateaus at discrete values as a function of initial $\delta$. 

Production, observation and control of anyons is one of the most intriguing quests motivating the considerable efforts towards FQHE regime. Anyons are quasiparticles with the peculiar property of satisfying non common braiding rules when moving around each other. In the Laughlin case the anyonic excitations are quasiholes $\prod_{i}(z_{i}-z_{\mathrm{qh}})\Psi_{\mathrm{Lau}}$, which can be produced and controlled by impinching a narrow strongly repulsive laser beam at position $z_{\mathrm{qh}}$ as put forward in \citep{paredes01}. A further feature of our proposal is that addressing a giant vortex with $\ell\ge N$ permits in principle to get a final state with a whole manifold of anyonic quasiholes and to study its exotic properties.

\section*{Acknowledgements}

We acknowledge fruitful discussions with T. Busch, J.I. Cirac, N.
Gemelke, M. Haque and\textbf{ }G. Juzeli$\bar{\mathrm{u}}$nas\textbf{.}
M. Rizzi has received funding from the European Community\textquoteright{}s
Seventh Framework Programme (FP7/2007-2013) under grant agreement
no. 247687 (IP-AQUTE). M. Roncaglia was partially supported by the EU-STREP
Projects HIP (Grant No. 221889) and COQUIT (Grant No. 233747). 
J.D. acknowledges support by IFRAF, ANR (BOFL project) and the European Union (MIDAS project).

\appendix

\section*{Methods}

\textbf{\small LLL approximation.}{\small{} Freezing the longitudinal
degrees of freedom by a large $\omega_{z}$, the single-particle Hamiltonian
(\ref{eq:Htrap}) can be rewritten in dimensionless units as\[
\mathcal{H}_{\mathrm{trap}}=(2a^{\dagger}a+1)+\delta(b^{\dagger}b-a^{\dagger}a),\]
with $a^{\dagger}=-\partial_{z}+\bar{z}/2$ and $b^{\dagger}=-\partial_{\bar{z}}+z/2$
in terms of $z=(x+iy)$. Eigenstates with angular momenta $l=m-p$
and energies $\epsilon_{m,p}=2p+l\delta$ are built by iteratively
applying the ladder operators $a^{\dagger}$ and $b^{\dagger}$ on
the vacuum $\psi_{0,0}\equiv\langle z\vert0,0\rangle=e^{-\vert z\vert^{2}/2}/\sqrt{\pi}$,
i.e. $\vert m,p\rangle=\frac{(a^{\dagger})^{p}(b^{\dagger})^{m}}{\sqrt{p!m!}}\vert0,0\rangle$.
The explicit wavefunctions are\[
\psi_{m,p}=\psi_{0,0}\cdot\frac{z^{m-p}}{\sqrt{p!m!}}\sum_{q=0}^{p}(-1)^{q}\binom{p}{q}\,\binom{m}{q}q!\vert z\vert^{2(p-q)},\]
which can be rewritten in terms of confluent hypergeometric functions
$U(-p,\, m-p+1,\,\vert z\vert^{2})$. Energy levels for small $\delta$'s
are organized in quasi-degenerate manifolds called Landau
Levels (LL), labelled by the integer $p$ and separated by an energy
gap $2$. When dealing with many-body problems, the usual approximation
is to cut down the single-particle Hilbert space to the lowest LL
(LLL) $p=0$ where wavefunctions (apart from Gaussian weight) are
analytical in $z$, being $\psi_{m,0}=\psi_{0,0}\cdot z^{m}/\sqrt{m!}$.
This is well justified and valid if the chemical potential remains
well under the LL gap value $2$; furthermore, the LL mixing due to
the interaction term in the strongly correlated FQHE regime is negligible \citep{morris06}. }{\small \par}

{\small Here we show that the LLL approximation remains valid even
in the presence of a plug laser (\ref{eq:Ur}). The matrix
elements within the LLL are\begin{equation}
\left\langle m,0\left|U_{w}(r)\right|m,0\right\rangle =\alpha\left(1+\frac{2}{w^{2}}\right)^{-(m+1)},\label{eq:U_LLL}\end{equation}
which provide the energy shifts due to $U_{w}$ at the  first order of perturbation theory.
The rotationally symmetric $U_{w}$ only couples states with the same
angular momentum and different LL labels:
\begin{eqnarray}
\left\langle m+p,p\left|U_{w}\right|m+p,p\right\rangle  & \simeq & \left\langle m,0\left|U_{w}\right|m,0\right\rangle \cdot\left(1+\frac{2}{w^{2}}\right)^{-p},\nonumber \\
\left\langle m,0\left|U_{w}\right|m+p,p\right\rangle  & \sim & (\sqrt{m}/w^{2})^{p}\,\left\langle m,0\left|U_{w}\right|m,0\right\rangle .\label{eq:mixingLL}\end{eqnarray}
Due to the exponential decay with $m$ in Eq.(\ref{eq:U_LLL}), the plug laser
affects mainly low angular momenta, localized inside a circular area
of radius $\sqrt{m}<w$. For large $w$ the energy shift of higher
LL is quite similar to the LLL one, thus almost preserving the distance
$2$ between adjacent LLs. The inter-LL terms in the illuminated region $m<w^{2}$ are
reduced by the prefator $\sqrt{m}w^{-2}$, varying exponentially with the LL
index $p$.   
Formally, one is allowed to use LLL approximation only for those $m$ such that the 
matrix elements (\ref{eq:mixingLL}) are $\ll2$. In practice among the exact eigenstates $\phi_{m,0}$ of $\mathcal{H}_{\mathrm{trap}}+U_{w}$, those that differ significantly from the unperturbed states $\psi_{m,0}$ have a very small occupation in the many-body solution: the global LL mixing
$\sum_m [ (1 - | \langle  \psi_{m,0} | \phi_{m,0} \rangle |^2)\langle  n_m \rangle] $
is bounded by $1\%$ for all the simulations presented
(see Fig.\ref{fig:potential3D}c). }{\small \par}

{\small \medskip{}
}\textbf{\small Numerics.}{\small{} LLL approximation reduces the particles'
degrees of freedom to one, the angular momentum $m$, and the many
body system is then described by the Fock basis $\left|n_{0}n_{1}\ldots\right\rangle $.
The Hilbert space available for a single particle is further cut to
$0\le m\le2N$ in order to realize numerics without affecting the
correct description of the Laughlin state and its lowest excitations;
occupation number in higher $m$'s never exceeds negligible amounts.
Even with this strong reduction, the dimension of the largest Hilbert
subspace considered for $N$ particles grows as $\sim1.75\cdot10^{0.74\, N-2}$  
($\sim 4\cdot10^5$ for $N=10$).
Within the LLL approximation, central contact interactions of the
form (\ref{eq:2-body delta potential}) could be written in terms
of a single Haldane pseudopotential \citep{Haldane83}:
\begin{equation}
\mathcal{H}_{2}=\frac{c_{2}}{4\pi}\sum_{m_{0}m_{1}m_{2}}\frac{1}{2^{m_{0}}}\sqrt{\binom{m_{0}}{m_{1}}\,\binom{m_{0}}{m_{2}}}d_{m_{0}-m_{1}}^{\dagger}d_{m_{1}}^{\dagger}d_{m_{2}}d_{m_{0}-m_{2}}\ .
\label{eq:2-body-2ndquant}
\end{equation}
Here we introduced the second quantization operator $d_{m,p}^{\dagger}$, which creates a particle in $\psi_{m,p}$, and used the simplified notation $d_{m,0}^{\dagger}\equiv d_m^\dagger$ for LLL states. The coefficients in (\ref{eq:2-body-2ndquant}) account for pairs having total angular momentum $m_{0}$ with null
component in the center of mass frame. This leads to a sparse matrix
form for the interactions, with an average filling per row growing
as $\sim0.2\cdot N^{2.92}$. For $N=10$ particles and $L=L_{\mathrm{Lau}}=90$,
we need $\sim1$ Gb RAM to store the Hamiltonians and less than one hour CPU-time 
on a single-core 3GHz desktop processor to diagonalize a single instance of the problem.}{\small \par}

{\small \medskip{}
}\textbf{\small Condensate in the ring.}{\small{} Given the noninteracting
energies (\ref{eq:E_m}) with a minimum in $m=\ell$, the GS of $\mathcal{H}_{1}=\sum_{m}\epsilon_{m}d_{m}^{\dagger}d_{m}$
is given by the giant vortex (\ref{eq:giant vortex}). 
The angular momentum $\ell$ gets depleted by the insertion of two body interactions
(\ref{eq:2-body-2ndquant}) in favour of the nearest ones $\ell+q,\,\ell-q$,
with $q\ll\ell$. For a condensate of $N_{0}$ particles in $m=\ell$,
the most dominant terms in $\mathcal{H}_{2}$ are\[
\mathcal{H}_{2}=\frac{c_{2}}{4\sqrt{\pi^{3}\ell}}\left[N_{0}^{2}+N_{0}\sum_{q\neq0}2\beta_{q}^{\dagger}\beta_{q}+2\beta_{-q}^{\dagger}\beta_{-q}+\beta_{l}\beta_{-q}+\beta_{-q}^{\dagger}\beta_{q}^{\dagger}\right],\]
where new operators $\beta_{q}=d_{\ell+q}$ have been defined and
the Stirling approximation $n!\approx\sqrt{2\pi}e^{-n}n^{n+\frac{1}{2}}$
employed. The coupling $c_{2}$ gets renormalized by $\ell^{-1/2}$
as a consequence of the wavefunction localization on a ring of length
$2\pi\sqrt{\ell}$. Eliminating $N_{0}$ by the number operator 
$\hat{N}=N_{0}+\sum_{q}\beta_{q}^{\dagger}\beta_{q}$,
the overall Hamiltonian $\mathcal{H}_{1}+\mathcal{H}_{2}$ reads 
\begin{eqnarray}
\mathcal{H} & = & N\epsilon_{\ell}+\frac{g}{2}N^{2}+\sum_{q}\left(\bar{\epsilon}_{q}+gN\right)\beta_{q}^{\dagger}\beta_{q}\nonumber \\
 &  & +\frac{g}{2}N\sum_{q}\left(\beta_{q}\beta_{-q}+\beta_{-q}^{\dagger}\beta_{q}^{\dagger}\right),\label{eq:H quad}\end{eqnarray}
with $g=\frac{c_{2}}{2\sqrt{\pi^{3}\ell}}$ and $\bar{\epsilon}_{q}\equiv\epsilon_{\ell+q}-\epsilon_{\ell}\approx2\alpha q^{2}/(e\ell)^{2}$
for small $q$ and $e$ is the exponential constant. Under the proposed
scaling $\alpha\propto\ell=N-1$, increasing $N$ enhances the importance
of interactions with respect to single-body energies. }{\small \par}

{\small The quadratic bosonic model (\ref{eq:H quad}) can be exactly
solved \citep{FetterWalecka71} by the Bogoliubov transformation 
$\beta_{q}=u_{q}\eta_{q}-v_{q}\eta_{-q}^{\dagger}$
with $u_{q}^{2}-v_{q}^{2}=1$, by which it reads \[
\mathcal{H}=N\epsilon_{\ell}+\frac{g}{2}N^{2}+\frac{1}{2}\sum_{q\neq0}\left(\Lambda_{q}-\bar{\epsilon}_{q}-gN\right)+\sum_{q\neq0}\Lambda_{q}\eta_{q}^{\dagger}\eta_{q}.\]
Due to the positiveness of $\Lambda_{q}$, the GS $|\Phi_{0}\rangle$
(an approximation to the exact GS $|\Psi_{0}\rangle$) is given by
the vacuum of quasiparticle excitations, $\eta_{q}|\Phi_{0}\rangle=0$,
$\forall q$.}{\small{} 
The quasiparticle spectrum is $\Lambda_{q}=\sqrt{\left(\bar{\epsilon}_{q}+gN\right)^{2}-g^{2}N^{2}}$
and the occupation number of the state $q$ is $v_{q}^{2}=\frac{1}{2}\left(\frac{\bar{\epsilon}_{q}+gN}{\Lambda_{q}}-1\right)$.
The depletion is the fraction of particles outside the condensate,
\begin{equation}
\frac{N-N_{0}}{N}=\frac{1}{N}\sum_{q\neq0}v_{q}^{2}=\frac{1}{2N}\sum_{q\neq0}\left[\frac{q^{2}+Q^{2}}{\sqrt{\left(q^{2}+Q^{2}\right)^{2}-Q^{4}}}-1\right]\label{eq:depletion}
\end{equation}
with $Q^{2}=gN(e\lambda)^{2}/2\alpha$. The expression for $v_{q}^{2}$
converges as $q^{-4}$ for high momenta, and no infrared divergencies
appear since angular momenta are quantized in integers. For large
$Q$, Eq.(\ref{eq:depletion}) becomes $1-N_{0}/N\approx Q\log Q/(\sqrt{2}N)$
which vanishes for $N\to\infty$ only if $Q\propto N^{1-\zeta}$,
with $\zeta>0$. In the same limit the GS energy turns out to be \[
E\approx N\,\left(\epsilon_{\ell}+\frac{gN}{2}-\frac{gQ}{3\sqrt{2}}+\frac{g}{4}\right)\]
 and the chemical potential 
 \begin{equation}
\mu=\frac{\partial E}{\partial N}\approx gN\left[1-\frac{e}{6}\left(\frac{gN}{\alpha}\right)^{1/2}\right],\label{eq:chem pot}
\end{equation}
whose leading term is the expected result from the Gross--Pitaevskii
approach in the LLL. Working at constant $\mu<2$, as required by
LLL approximation, implies a scaling $c_{2}=C\, N^{-1/2}$, with $C=2\sqrt{\pi^{3}}\mu$
(see Fig.\ref{fig:Finite-size-scaling}). Moreover, $\alpha\propto N$
implies that $Q\propto N^{1/2}$, ensuring the vanishing of the depletion
fraction. The radial confinement is rather strong, since the standard deviation
$\Delta q$ grows only as $N^{1/4}$, as deduced from the calculation
$(\Delta q)^{2}=\sum_{q}q^{2}v_{q}^{2}/N\approx\sqrt{2}Q^{3}/(3N)$. }{\small \par}

{\small The energy gap to the first excitation }$|\Phi_{1}\rangle=\eta_{1}^{\dagger}\eta_{-1}^{\dagger}|\Phi_{0}\rangle${\small{}
with the same $L$ reads \[
\Delta=\Lambda_{1}+\Lambda_{-1}=2gN\sqrt{(Q^{-2}+1)^{2}-1}\approx\frac{2\sqrt{2}\mu}{Q}\]
and vanishes as $N^{-1/2}$ just as $c_{2}$, i.e. the energy scale
of the final Laughlin state (Fig.\ref{fig:Finite-size-scaling}).
Within such Bogoliubov analysis, it is also possible to determine
the scaling of many other interesting quantities: e.g. the interaction
energy $\langle\mathcal{H}_{2}\rangle=c_{2}\partial E/\partial c_{2}$
scales as $N$.}{\small \par}

{\small \medskip{}
}\textbf{\small Robustness against trap defects.} {\small The main experimental defects that may hinder our protocol are a residual static quadrupole anisotropy $u$ and  an off-centering $v$ of the plug beam, which are described by the single-particle potentials $\mathcal{H}_u=u(x^{2}-y^{2})/2$ and $\mathcal{H}_v=\alpha\exp\left[-2[(x-v)^{2}+y^{2}]/w^{2}\right]-U_w$, respectively.  Both terms break the rotation symmetry and couple manifolds corresponding to different total angular momenta.  We consider first the coupling $\mathcal{H}_{u} $ since it turns out to have the largest impact for practical conditions. Its second-quantized expression can be written $\mathcal{H}_{u}=\mathcal{H}_{u}^{(0)}+\mathcal{H}_{u}^{(1)}+\mathcal{H}_{u}^{(2)}$ with
\begin{eqnarray*}
\mathcal{H}_{u}^{(0)}&=&\frac{u}{4}\sum_{m,p}\sqrt{(m+1)(m+2)}\,(d_{m+2,p}^{\dagger}d_{m,p}+\mbox{H.c.}) \ , \\
\mathcal{H}_{u}^{(1)}&=&\frac{u}{2}\sum_{m,p}\sqrt{m(p+1)}\,(d_{m-1,p+1}^{\dagger}d_{m,p}+\mbox{H.c.})
\ , \\ 
\mathcal{H}_{u}^{(2)}&=&\frac{u}{4}\sum_{m,p}\sqrt{(p+1)(p+2)}\,(d_{m,p+2}^{\dagger}d_{m,p}+\mbox{H.c.}) \ ,
\end{eqnarray*}
where $d_{m,p}^{\dagger}$ creates a particle in the one-body state $\psi_{m,p}$.  In the absence of trap defects, we assume that the adiabatic approximation is valid and that the system is at any time in the instantaneous  GS $|\Psi_{0}\rangle$ of the time-dependent Hamiltonian $\mathcal{H}_{0}$, given by the trapping potential plus the interaction energy. The state $|\Psi_{0}\rangle$ has angular momentum $L$, energy $E_0$ and belongs to the LLL ($p=0$). We estimate the influence of $\mathcal{H}_u$ by calculating perturbatively  the depletion of the probability for finding the system in $|\Psi_{0}\rangle$.  The states that can be reached are given at first order by the action of the various terms in $\mathcal{H}_{u}$ on $|\Psi_{0}\rangle$. The first contribution $\mathcal{H}_{u}^{(0)}$ induces a coupling to other LLL states having angular momentum $L\pm 2$. The contributions $\mathcal{H}_{u}^{(1)}$ and $\mathcal{H}_{u}^{(2)}$ connect $|\Psi_{0}\rangle$ with states having both angular momentum $L-2$ and belonging to the first and second excited LL, respectively. We denote these normalized states as  $|\Psi_{-2}^{1LL}\rangle \propto \mathcal{H}_{u}^{(1)}|\Psi_{0}\rangle$ and $|\Psi_{-2}^{2LL}\rangle \propto \mathcal{H}_{u}^{(2)}|\Psi_{0}\rangle$. 

In analogy with the textbook problem of Rabi oscillations, we find that the initial state $|\Psi_{0}\rangle$ is protected from depletion towards another eigenstate $|\Psi_{\alpha}\rangle$ of $\mathcal{H}_{0}$ with energy $E_\alpha$
if  the energy detuning  $|E_{\alpha}-E_0|$ is much bigger than the coupling $\Gamma_{\alpha}=|\langle\Psi_{\alpha}|\mathcal{H}_{u}|\Psi_{0}\rangle|$. In the opposite case where $\Gamma_{\alpha}\gg |E_{\alpha}-E_0|$,  the population of $|\Psi_0\rangle$ slims down as $1-\sin^{2}\left(\Gamma_{\alpha} t\right)$. 
In such a resonant case,  we should have $\Gamma_{\alpha}\ll T^{-1}$ where $T$ is the total time of evolution, to avoid any significant leakage from the initial state $|\Psi_{0}\rangle$. Once the stirring has been stopped, the dominant  Hamiltonian \emph{in the laboratory frame} corresponds to the single-particle motion in the trapping potential
\[
\mathcal{H}_\mathrm{trap}^{(0)}=\sum_{m,p}(m+p)d_{m,p}^{\dagger}d_{m,p}\ .
\]
With respect to $\mathcal{H}_\mathrm{trap}^{(0)}$, the state $|\Psi_{-2}^{1LL}\rangle$ has the same energy as $|\Psi_{0}\rangle$ and the coupling between these two states is thus the dominant escape route from  $|\Psi_{0}\rangle$. To estimate the corresponding rate, we concentrate on the  last part of the adiabatic evolution and we take $|\Psi_{0}\rangle$ equal to the Laughlin state, where $L=N(N-1)$. The coupling matrix element is then
\[
\Gamma_{-2}^{1LL}=|\langle\Psi_{-2}^{1LL}|\mathcal{H}_{u}|\Psi_{0}\rangle|=\frac{u}{2}\sqrt{\sum_{m}m\langle\Psi_{0}|n_{m,0}|\Psi_{0}\rangle}=\frac{u}{2}\sqrt{L}\approx \frac{u}{2}N\ .
\]
Since $|\Psi_{-2}^{1LL}\rangle$ and $|\Psi_{0}\rangle$ have the same trapping energy, the detuning $|E_{-2}^{1LL}-E_0|$ originates solely from the difference in interaction energy. More precisely a lower bound for this detuning is the Laughlin gap $\Delta_{\rm Lau}\approx0.1c_{2}$. 
Hence, for $N=10$ and $c_{2}=0.5$, the non-resonant condition $\Gamma_{-2}^{1LL}\ll |E_{-2}^{1LL}-E_0|$ is satisfied if the defect amplitude $u$ is much less than $10^{-2}$. For $u=10^{-2}$, we expect the population in $|\Psi_{0}\rangle$ to decay in a time on the order of $1/\Gamma_{-2}^{1LL}\sim 20$. 

The off-centering defect $\mathcal{H}_v$ expanded at first order in $v$ connects $|\Psi_{0}\rangle$ only with states $|\Psi_{\alpha}\rangle$ whose energy detuning is equal to $1$. This large detuning is favourable to minimise the departure rate from $|\Psi_0\rangle$. Moreover, the influence of this defect fades away together with the plug during time evolution. Hence, repeating a similar analysis as for $\mathcal{H}_u$, we eventually find that the condition $\Gamma_{\alpha}\ll | E_{\alpha}-E_0|$ is safely fulfilled when $v<1$ (in units of the trap length), which is not a very stringent condition in practice. }

\bibliographystyle{naturemag}
\bibliography{hall}

\end{document}